\documentclass[
]{ceurart}

\usepackage{listings}
\usepackage{xcolor}
\usepackage{paralist}
\usepackage[inline]{enumitem}%
\usepackage{fancyvrb}
\usepackage{subcaption}
\definecolor{codegreen}{rgb}{0,0.6,0}
\definecolor{codegray}{rgb}{0.5,0.5,0.5}
\definecolor{codepurple}{rgb}{0.58,0,0.82}
\definecolor{backcolour}{rgb}{0.95,0.95,0.92}
\lstdefinestyle{mystyle}{
    backgroundcolor=\color{backcolour},   
    commentstyle=\color{codegreen},
    keywordstyle=\color{black},
    numberstyle=\tiny\color{codegray},
    stringstyle=\color{codepurple},
    basicstyle=\ttfamily\footnotesize,
    breakatwhitespace=false,         
    breaklines=true,                 
    captionpos=b,                    
    keepspaces=true,                 
    numbers=left,                    
    numbersep=5pt,                  
    showspaces=false,                
    showstringspaces=false,
    showtabs=false,                  
    tabsize=2
}
\begin{document}

\copyrightyear{2026}
\copyrightclause{Copyright for this paper by its authors.
  Use permitted under Creative Commons License Attribution 4.0
  International (CC BY 4.0).}

\conference{Workshop on LLM-driven Knowledge Graph and Ontology Engineering (llms4kgoe) 2026 co-located with ESWC 2026}



\title{Benchmarking Resource-Efficient LLMs for Research Topic Ontology Generation in the Biomedical Field}


\author[1]{Tanay Aggarwal}[%
orcid=0009-0009-9477-7112,
email=tanay.aggarwal@open.ac.uk,
]
\cormark[1]
\address[1]{Knowledge Media Institute, The Open University, Milton Keynes, UK}

\author[1]{Angelo Salatino}[%
orcid= 0000-0002-4763-3943,
email=angelo.salatino@open.ac.uk,
]

\author[1,2]{Francesco Osborne}[%
orcid=0000-0001-6557-3131,
email=francesco.osborne@open.ac.uk,
]
\address[2]{Department of Business and Law, University of Milano-Bicocca, Milan, IT}

\author[1]{Enrico Motta}[%
orcid=0000-0003-0015-1952,
email=enrico.motta@open.ac.uk,
]

\cortext[1]{Corresponding author.}

\begin{abstract}
    Knowledge Organization Systems like Ontologies and taxonomies are fundamental for structuring scientific knowledge, yet their manual curation presents a persistent bottleneck in knowledge management. While Large Language Models (LLMs) offer a scalable mechanism for automated ontology generation, their capacity to classify complex, domain-specific semantics requires systematic evaluation. In this paper, we assess the performance of five small, open-source LLMs (up to 9 billion parameters) in identifying semantic relationships between biomedical concepts. To support this evaluation, we introduce \textit{MeSH-Rel-4K}, a dataset comprising 4K semantic relationships extracted from the Medical Subject Headings (MeSH). We analyse three adaptation strategies: standard prompting, Chain-of-Thought prompting, and fine-tuning. 
    While parameter-constrained models traditionally struggle with the nuances of in-context logic, our results reveal that targeted fine-tuning increases the average F1-score by 34.1 percentage points. 
    %
    %
    %
    These results confirm that direct fine-tuning effectively exceeds the reasoning bottlenecks of smaller LLMs, providing an accurate, automated methodology for the construction and evolution of specialised biomedical ontologies.
    %
\end{abstract}


\maketitle

\section{Introduction}\label{sec:introduction}

Knowledge Organization Systems (KOSs), such as ontologies and taxonomies, have emerged as essential frameworks for systematically structuring and interlinking information. Within the research ecosystem, these systems are particularly critical for mapping topics, thereby enhancing data retrievability and management in digital libraries~\cite{salatino2025survey}.
These systems are utilised by major publishers like Springer Nature\footnote{The SpringerNature Taxonomy to organise their published content}, IEEE\footnote{The IEEE Theasurus is used to classify academic articles and research within the IEEE digital library}, and PubMed\footnote{Medical Subject Headings (MeSH) is used to organise academic content on PubMed}~\cite{Dunne2020,lipscomb2000medical,rous2012major} to categorise research outputs, e.g., research articles, books, monographs. Further, KOSs support smart downstream applications to navigate and interpret academic literature, such as recommender systems~\cite{palumbo2020entity2rec}, classifiers~\cite{cadeddu2024comparative}, search engines~\cite{gusenbauer2020academic}, conversational agents~\cite{meloni2023integrating}, and analytics dashboards~\cite{angioni2021aida}.

However, as shown by Salatino et al.~\cite{salatino2025survey} existing ontologies of research areas exhibit uneven coverage and are updated infrequently, causing them to lag behind rapidly emerging research areas. These limitations stem from the creation and maintenance of ontologies that require coordinated manual effort from domain experts over prolonged periods~\cite{osborne2015klink}. Previous approaches to automate ontology generation encounter structural limitations when processing the granular complexity of scientific terminologies~\cite{han2020wikicssh, osborne2012mining, osborne2015klink, sanderson1999deriving}. Consequently, the development of domain-specific ontologies remains predominantly a manual process.

Recent advances in Large Language Models (LLMs) present alternative methodologies for ontology generation~\cite{babaei2023llms4ol, aggarwal2026large, lippolis2025ontology}. In this paper, we analyse the performance of five small, open-source LLMs (up to 9 billion parameters) in identifying semantic relations between pairs of biomedical research topics. We test three adaptation strategies: standard prompting, Chain-of-Thought (CoT) prompting~\cite{wei2022chain}, and parameter-efficient fine-tuning~\cite{peft}. Our analysis tests the capacity of LLMs to identify three explicit semantic relations: \texttt{broader}, \texttt{narrower}, and \texttt{same-as}, alongside an \texttt{other} category for unrelated concepts.

To support this investigation, we extracted 4K semantic relationships from MeSH, introducing \textit{MeSH-Rel-4K}. Our experiments demonstrate that while CoT prompting yields moderate baseline improvements, explicit fine-tuning systematically increases classification accuracy. The fine-tuned \texttt{gemma-2-9b-it} achieved an F1-score of 91.6\%, while the smallest model (\texttt{Llama-3.2-3B-Instruct}) recorded a 60.5 percentage point increase in its F1-score post-fine-tuning.

We release our complete codebase and the \textit{MeSH-Rel-4K} dataset in our GitHub repository\footnote{Our code and dataset - \url{https://github.com/ImTanay/Biomedical-Ontology-Generation}}.
%



The rest of the paper is structured as follows: Section~\ref{related-work} discusses the related literature. Section~\ref{background} defines the task under investigation, describes the \textit{MeSH-Rel-4K} dataset, and details the selected models. Section~\ref{methodology} outlines the experimental methodology and implementation. Section~\ref{results} presents the results and an error analysis. Finally, Section~\ref{conclusion} concludes this study, outlining areas for future research.


\section{Related Work}\label{related-work}
This section discusses the literature focusing on two key aspects relevant to our work: ontologies of research areas and state-of-the-art methodologies for generating these ontologies.

\subsection{Ontologies of Research Areas}\label{ontologies-of-research-areas}

Ontologies of research areas are essential for organising and retrieving research outputs, such as publications and datasets, within digital libraries and repositories~\cite{zeng2008knowledge}.  By delineating subfields and their semantic relationships, ontologies provide structured maps of academic domains. 
Notable examples include domain-specific ontologies, such as Medical Subject Headings (MeSH)~\cite{lipscomb2000medical}, 
 Systematized Nomenclature of Medicine - Clinical Terms (SNOMED CT)~\cite{lee2014literature}, and the Gene Ontology (GO)~\cite{ashburner2000gene} within the biomedical domain. Other academic disciplines rely on resources such as
the ACM Computing Classification System (CCS)~\cite{rous2012major}, the Computer Science Ontology (CSO)~\cite{salatino2018computer}, the IEEE Thesaurus\footnote{IEEE Thesaurus - \url{https://www.ieee.org/content/dam/ieee-org/ieee/web/org/pubs/ieee-thesaurus.pdf}}, the Mathematical Subject Classification (MSC), and Physics Subject Headings (PhySH)~\cite{smith2020physics}.
In Biomedicine, MeSH is a comprehensive ontology comprising over 30K concepts, developed and maintained by the National Library of Medicine~\cite{lipscomb2000medical}. Widely adopted across the medical and health sciences, it is updated annually.
On the other hand, SNOMED CT encompasses over 376K distinct medical concepts, functions as a highly granular clinical terminology for electronic health records to standardise clinical documentation and reporting~\cite{chang2021use}.
Finally, the Gene Ontology includes more than 43K concepts to classify molecular functions, biological processes, and cellular components across various species and genomic databases~\cite{gene2019gene}.
%
%
%
In Computer Science, the ACM CCS\footnote{The ACM CCS – \url{http://www.acm.org/publications/class-2012}} is a taxonomy of research topics within the field of computer science, encompassing approximately 2K research topics. Curated by the Association for Computing Machinery (ACM), its most recent update occurred in 2012.
The IEEE Thesaurus primarily covers the field of engineering while incorporating various concepts relevant to computer science. It is developed by the Institute of Electrical and Electronics Engineers\footnote{IEEE Taxonomy - \url{https://www.ieee.org/content/dam/ieee-org/ieee/web/org/pubs/ieee-taxonomy.pdf}}, contains 5.6K concepts and 24K semantic relationships, and is updated annually.
The MSC is a taxonomy with more than 6.5K topics, encompassing a wide range of mathematical disciplines, from pure mathematics to statistics, jointly curated by the American Mathematical Society and zbMATH. It is updated decennially~\cite{Dunne2020}.
%
%
PhySH features approximately 3.7K concepts related to physics and is primarily employed to index literature in Physical Review and on arXiv~\cite{smith2020physics}. Maintained by the American Physical Society, it is updated annually, with the most recent release occurring in December 2025.

Furthermore, multi-disciplinary ontologies play a crucial role across broader academic landscapes. The UNESCO Thesaurus, curated by UNESCO, encompasses 4.4K subjects and is predominantly utilised for indexing and retrieving resources within UNESCO’s document repository. It undergoes frequent minor revisions, typically every two to three months.
Similarly, the ANZSRC Fields of Research (FoR) was developed by the Australian and New Zealand research councils. It covers 4.4K topics across diverse disciplines, though it is updated less frequently ($\sim$10 years). Digital Science employs this taxonomy to organise research content within its primary datasets, Dimensions.ai and Figshare.
%
%
OpenAlex Topics is another multi-disciplinary taxonomy comprising 4.8K topics. It is primarily utilised within OpenAlex, a free and open catalogue of scholarly articles, and is currently in its initial release~\cite{openalex2024}.

A systematic survey~\cite{salatino2025survey} of 45 widely adopted ontologies that describe research areas revealed that none of them is simultaneously openly available, sufficiently fine-grained, comprehensively describes all academic disciplines, and is frequently maintained.
Part of this gap is due to the high costs of manual curation. Indeed, 82\% of them are manually curated, demanding substantial time and financial resources. We argue that automated frameworks have emerged as a highly viable alternative to bypass these resource bottlenecks and ensure the sustainable evolution of scientific knowledge bases.

\subsection{Automatic Generation of Research Area Ontologies}\label{automatic-generation}

Early semi-automatic ontology generation progressed from combining expert input with statistical tools~\cite{maedche2001learning} to utilising machine learning frameworks (e.g., Text2Onto~\cite{cimiano2005text2onto}, OntoLearn~\cite{velardi2013ontolearn}) to minimise manual effort. More recently, deep learning models such as BERT have significantly advanced concept extraction and hierarchical relationship modelling~\cite{chen2020constructing, devlin2019bert, grootendorst2022bertopic}.

LLMs have further demonstrated strong potential for ontology generation~\cite{babaei2023llms4ol}. However, persistent accuracy and consistency issues dictate that frameworks such as NeOn-GPT~\cite{fathallah2024neon} and LLMs4Life~\cite{fathallah2024llms4lifelargelanguagemodels} function best as assistive tools requiring context-rich prompting and human-in-the-loop verification~\cite{lippolis2025ontology, saeedizade2024navigating, tsaneva2025knowledge}. Consequently, LLMs currently struggle to autonomously produce expert-level ontologies for specialised tasks like scientific publication classification~\cite{sun2024large}. 

Specifically within the domain of research topic ontologies, prior semi-automated and automated approaches have successfully employed subsumption logic, co-occurrence data, and citation clustering to construct large-scale frameworks. Notable examples include the CSO~\cite{salatino2018computer, osborne2015klink}, Microsoft’s Field of Science~\cite{wang2020microsoft}, and taxonomies for OpenAlex~\cite{openalex2024} and 
an ANZSRC extension~\cite{jenset2025large}, alongside methods designed to enrich existing ontologies~\cite{kotis2020ontology, osborne2018pragmatic}.

Although LLMs are increasingly utilised for broader academic tasks, including paper discovery~\cite{chow2024semantic}, citation prediction~\cite{buscaldi2024citation}, scientific question answering~\cite{auer2023sciqa}, and literature review generation~\cite{bolanos2024artificial}, their specific application in generating research topic ontologies remains underexplored. This study addresses this gap by systematically evaluating the capacity of recent LLMs to infer semantic relationships between research topics.

\section{Background}\label{background}

This section outlines the task under investigation (Section~\ref{task}), details the \textit{MeSH-Rel-4K} dataset (Section~\ref{mesh-rel-4k}), and provides an overview of LLMs utilised in this study (Section~\ref{llms}).

\subsection{Task Definition}\label{task}

The task is defined as identifying the semantic relationship between a pair of research topics ($t_A$, $t_B$). It is formulated as a single-label, multi-class classification problem, where each pair is assigned to one of the following four categories:
\begin{itemize}\itemsep=-4pt
    \item \texttt{broader}: $t_A$ is a parent topic of $t_B$. For example, \textit{plants} is broader than \textit{agricultural crops}. 
    \item \texttt{narrower}: $t_A$ is a child topic of $t_B$. For example, \textit{pneumocystis infections} is contained within \textit{fungal diseases}. This constitutes the inverse relationship to \texttt{broader}.
    \item \texttt{same-as}: $t_A$ and $t_B$ can be used interchangeably to represent the same research concept. For example, \textit{taste dysfunction} and \textit{taste disorders}.
    \item \texttt{other}: $t_A$ and $t_B$ are unrelated or when none of the above applies. In contrast to the previous three categories, this does not define a semantic relationship. It rather provides a mechanism for the classifier to label negative examples.
\end{itemize}

In alignment with our previous study~\cite{aggarwal2026large}, we considered only \texttt{broader}, \texttt{narrower}, and \texttt{same-as} as semantic relationships, since these are fundamental for constructing ontologies that effectively map academic disciplines.

\subsection{The MeSH-Rel-4K Dataset}\label{mesh-rel-4k}

To evaluate the efficacy of fine-tuned LLMs in inferring semantic relationships between research topics compared to different prompting strategies, we sampled 4,000 semantic relationships from MeSH\footnote{Medical Subject Headings (MeSH) - \url{https://id.nlm.nih.gov/mesh/}}.  MeSH employs a custom schema\footnote{MeSH Schema - \url{http://id.nlm.nih.gov/mesh/vocab}} (\texttt{mesh:}), wherein each topic (\texttt{mesh:TopicalDescriptor}) is delineated by a set of properties. These topics are organised hierarchically via the \texttt{mesh:broaderDescriptor} property, while synonymous concepts are denoted using the \texttt{mesh:relatedConcept} property.
Utilising the January 2025 release of MeSH, we randomly sampled 1,000 relationship pairs for both the \texttt{broader} and \texttt{narrower} categories, deriving these from the \texttt{mesh:broaderDescriptor} property. Furthermore, we extracted 1,000 \texttt{same-as} instances based on the \texttt{mesh:relatedConcept} property, alongside 1,000 pairs 
%
%
%
of semantically disjoint topics to populate the \texttt{other} category.
%
%
%
%
The \textit{MeSH-Rel-4K} dataset was subsequently partitioned into training (2,800 semantic relationships), validation (400), and test (800) subsets, adhering to a 7:1:2 ratio. Moreover, research concepts were strictly isolated within their respective sets to prevent node leakage.
The complete dataset and the source code used for its construction are openly available in our GitHub repository\footnote{Our code - \url{https://github.com/ImTanay/Biomedical-Ontology-Generation}}.

\subsection{Large Language Models}\label{llms}

Building upon insights from our study in the engineering domain~\cite{aggarwal2026large}, we fine-tuned a selection of five high-performing, open-source LLMs: \texttt{mistral-7b}~\cite{jiang2023mistral}, \texttt{llama-3b}~\cite{llama3modelcard}, \texttt{gemma-9b}~\cite{team2024gemma}, \texttt{phi-3}~\cite{abdin2024phi3technicalreporthighly}, and \texttt{zephyr-7b}~\cite{huggingfaceUnslothzephyrsftHugging}. All models are 4-bit quantised and are available on HuggingFace.
%
Table~\ref{table:llms} provides an overview of these models, detailing their full names, the abbreviated aliases adopted throughout this paper, parameters count, context window sizes, and the specific Low-Rank Adaptation (LoRA)~\cite{hu2021lora} 
hyperparameters applied during fine-tuning: 
the rank of the update matrices (\textbf{r}\footnote{\textbf{r} - The rank of the update matrices (where lower values yield smaller update matrices with fewer trainable parameters)}) and the LoRA scaling factor (\textbf{alpha}). 
Initial selections for \textbf{r} and \textbf{alpha} followed the recommendations established by \cite{unslothLoRAFinetuning}. We then conducted multiple training iterations to empirically adjust these hyperparameters, and Table~\ref{table:llms} presents only the optimal configurations for each model.
%
%
%
\begin{table}[!h]
\centering
\caption{Overview of the five LLMs used in this study. The table includes the \textbf{Model} name, the \textbf{Alias} adopted in this paper, the trainable \textbf{Parameters} count, the size of context \textbf{Window}, the rank (\textbf{r}) and scaling factor (\textbf{alpha}) employed in LoRA.}
\label{table:llms}
\begin{tabular}{l|l|r|r|r|r}
\toprule
\textbf{Model}                       & \textbf{Alias}         & \textbf{Parameters} & \textbf{Window} & \textbf{r}   & \textbf{alpha} \\ \midrule
mistral-7b-instruct-v0.3    & mistral-7b     & 7.25B      & 32K    & 16  & 16    \\
Llama-3.2-3B-Instruct       & llama-3b      & 3.21B      & 128K   & 256 & 128   \\
gemma-2-9b-it               & gemma-9b      & 9.24B      & 8K   & 256 & 128   \\
Phi-3.5-mini-instruct       & phi-3         & 3.82B      & 128K   & 256 & 128   \\
zephyr-sft                  & zephyr-7b        & 7.24B      & 8K     & 16  & 16   \\
\bottomrule
\end{tabular}
\end{table}

\section{Methodology}\label{methodology}

This section details the procedures employed for the prompting (Section~\ref{prompting}) and fine-tuning (Section~\ref{fine-tuning}) experiments, alongside the technical specifications (Section~\ref{experiment}).

\subsection{Prompting Strategies}\label{prompting}

We utilised two distinct prompting techniques: standard and CoT prompting~\cite{wei2022chain,kojima2023largelanguagemodelszeroshot}.
%
%
In \textbf{standard prompting}, prompts are generated for each pair of research topics using a predefined template\footnote{Our prompt template is available at: \url{https://github.com/ImTanay/Biomedical-Ontology-Generation}}.
%
This template outlines the task, defines the four target relationships (as detailed in Section~\ref{task}), and instructs the LLMs to produce outputs in a structured format to facilitate automated parsing.

The \textbf{CoT, two-way prompting} strategy builds upon our previous findings~\cite{aggarwal2026large, aggarwal2024identifying}, which demonstrate the efficacy of this method in classifying semantic relationships. This approach employs a two-stage sequential prompting mechanism. The initial prompt instructs the model to define both topics, construct a sentence incorporating them, and reflect upon their potential semantic relationship. The subsequent prompt, in a chain-of-thought fashion, concatenates this generated output with specific classification instructions. To ensure robustness, the two topics are then swapped, and the entire process is repeated. Finally, a predefined set of empirical rules is applied to resolve any discrepancies between the bidirectional outcomes.
To ensure a fair and consistent comparison, identical prompt templates were administered across all evaluated models. Comprehensive details regarding the prompt templates and the empirical referee rules are described in~\cite{aggarwal2026large}.

\subsection{Fine-Tuning}\label{fine-tuning}

We fine-tuned five open-source LLMs on the \textit{MeSH-Rel-4K}, employing identical training and validation sets (as detailed in Section~\ref{mesh-rel-4k}) to ensure comparability across results and to maintain a standardised, prompt-based fine-tuning methodology. The base prompt template was structured as follows:

\begin{lstlisting}[language=Python]
Classify the relationship between '[TOPIC-A]' and '[TOPIC-B]'
\end{lstlisting}

Within this template, \texttt{'[TOPIC-A]'} and \texttt{'[TOPIC-B]'} serve as placeholders, which were dynamically replaced with specific pairs of research topics during the fine-tuning process. For instance:

\begin{lstlisting}[language=Python]
user: Classify the relationship between 'acinic cells' and 'cells'
model: relationship: 'narrower'
\end{lstlisting}

During both training and validation, the expected output was explicitly formulated as \texttt{relationship: [RELATIONSHIP-TYPE]}. This formatting conditioned the models to generate structured, easily parsable responses, thereby facilitating the automated extraction of the predicted relationships during evaluation.

\subsection{Experimental Setup}\label{experiment}

To fine-tune and interact with the LLMs detailed in Table~\ref{table:llms}, we utilised two open-source libraries: KoboldAI\footnote{KoboldAI - \label{kobold}\url{https://github.com/KoboldAI/KoboldAI-Client}} and Unsloth~\cite{unsloth}.

\textbf{KoboldAI} was employed to interact with the LLMs within a zero-shot prompting configuration. Built upon llama.cpp\footnote{llama.cpp - \url{https://github.com/ggml-org/llama.cpp}}, this platform facilitates API-based access to locally hosted models.

\textbf{Unsloth} was utilised for the fine-tuning process. This library is constructed upon the HuggingFace Transformers~\cite{wolf-etal-2020-transformers}, and Parameter-Efficient Fine-Tuning (PEFT)~\cite{peft} frameworks. It supports 4-bit quantised training via BitsAndBytes\footnote{BitsAndBytes - \url{https://github.com/bitsandbytes-foundation/bitsandbytes}}, substantially reducing memory consumption and permitting training on consumer-grade GPUs. Furthermore, Unsloth integrates LoRA~\cite{hu2021lora},  enabling PEFT by injecting lightweight, trainable adapter layers into the frozen pre-trained model architecture.

All experiments were executed within Google Colaboratory environments (using NVIDIA A100 and L4 GPUs). To facilitate reproducibility,
the complete codebase is openly accessible in our GitHub repository\footnote{Our code - \url{https://github.com/ImTanay/Biomedical-Ontology-Generation}}. This repository contains all necessary scripts for both the prompting and fine-tuning phases, alongside the precise configuration parameters applied within Unsloth and KoboldAI.

\section{Results and Discussion}\label{results}

This section presents the results of our experiments, beginning with the prompting strategies (Section~\ref{results-prompting}), followed by the outcomes of the fine-tuning (Section~\ref{results-fine-tuning}), and concluding with a comprehensive analysis of all LLMs fine-tuned on the \textit{MeSH-Rel-4K} (Section~\ref{results-mesh-4k}).

\subsection{Prompting Strategies}\label{results-prompting}
We evaluated the LLMs using precision, recall, and F1-score. Table~\ref{table:MeSH-Vanilla-STD} reports the results for standard prompting and Table~\ref{table:MeSH-Vanilla-CoT} for the CoT, two-way approach.

When comparing standard prompting against the CoT, two-way approach. The CoT, two-way approach shows consistently better performance, yielding an average F1-score increase of 5.4 percentage points (standard deviation $\pm 3.5$).
\texttt{gemma-9b} recorded the highest absolute metrics across both configurations (F1: 71.6\% with CoT, two-way and F1: 66.9\% with standard prompting). The largest relative gains occurred in \texttt{mistral-7b} (F1 improved by 10.0 percentage points) and \texttt{zephyr-7b} (F1 improved by 7.1 percentage points). However, the lowest-parameter model (\texttt{llama-3b}) registered a minimal 0.5 percentage point increase.
At the category level, metrics indicate that under standard prompting, recall for \texttt{same-as} and \texttt{narrower} falls strictly below that of \texttt{broader} and \texttt{other}. The CoT, two-way approach resolves these specific boundary deficits. For instance, \texttt{gemma-9b} recorded the highest overall recall (72.5\%) and precision (97.7\%) for \texttt{same-as}. \texttt{mistral-7b} also registered precision increases for both \texttt{broader} (76.6\%) and \texttt{narrower} (66.5\%).
These data confirm that structured reasoning improves the performance of LLMs, particularly for smaller and cost-effective models.

\begin{table}[!h]
\centering
\caption{Precision, Recall, and F1-score for standard prompting on the test set (800 semantic relationships) of \textit{MeSH-Rel-4K}. BR refers to performance on \texttt{broader} relations, NA to \texttt{narrower}, OT to \texttt{other}, and SA to \texttt{same-as}. AVG indicates the average performance across the four categories. The best-performing scores for each metric are highlighted in \textbf{\underline{bold \& underlined}}. Due to space constraints, the leading zero has been omitted from all values.}
\label{table:MeSH-Vanilla-STD}
\resizebox{\columnwidth}{!}{%
\begin{tabular}{l|ccccc|ccccc|ccccc}
\toprule
\multicolumn{1}{c|}{\multirow{2}{*}{\textbf{MODEL}}} & \multicolumn{5}{c|}{\textbf{F1-SCORE}} & \multicolumn{5}{c|}{\textbf{PRECISION}} & \multicolumn{5}{c}{\textbf{RECALL}} \\
\multicolumn{1}{c|}{} & AVG & BR & NR & OT & SA & AVG & BR & NR & OT & SA & AVG & BR & NR & OT & SA \\
\midrule
mistral-7b & .603 & \textbf{\underline{.706}} & .568 & .686 & .452 & .684 & \textbf{\underline{.692}} & .641 & .532 & .871 & .625 & .720 & .510 & \textbf{\underline{.965}} & .305 \\
llama-3b & .252 & .398 & .137 & .417 & .057 & .413 & .265 & .244 & .598 & .545 & .311 & .800 & .095 & .320 & .030 \\
gemma-9b & \textbf{\underline{.669}} & .705 & \textbf{\underline{.722}} & \textbf{\underline{.811}} & .438 & \textbf{\underline{.766}} & .571 & \textbf{\underline{.705}} & .788 & \textbf{\underline{1.000}} & \textbf{\underline{.694}} & \textbf{\underline{.920}} & \textbf{\underline{.740}} & .835 & .280 \\
phi-3 & .529 & .578 & .559 & .718 & .263 & .651 & .480 & .468 & \textbf{\underline{.794}} & .861 & .557 & .725 & .695 & .655 & .155 \\
zephyr-7b & .415 & .452 & .245 & .491 & \textbf{\underline{.472}} & .476 & .538 & .442 & .359 & .566 & .436 & .390 & .170 & .780 & \textbf{\underline{.405}} \\
\midrule
AVG & .494 & .568 & .446 & \textbf{\underline{.625}} & .336 & .598 & .509 & .500 & .614 & \textbf{\underline{.769}} & .525 & \textbf{\underline{.711}} & .442 & \textbf{\underline{.711}} & .235 \\
\bottomrule
\end{tabular}%
}
\end{table}

\begin{table}[!h]
\centering
\caption{Precision, Recall, and F1-score for CoT, two-way on Testset (800 semantic relationships) of \textit{MeSH-Rel-4K}. BR refers to performance on \texttt{broader} relations, NA to \texttt{narrower}, OT to \texttt{other}, and SA to \texttt{same-as}. AVG indicates the average performance across the four categories. The best-performing scores for each relation are highlighted in {\textbf{\underline{bold \& underlined}}}. Due to space constraints, the leading zero has been omitted from all values.}
\label{table:MeSH-Vanilla-CoT}
\resizebox{\columnwidth}{!}{%
\begin{tabular}{l|ccccc|ccccc|ccccc}
\toprule
\multicolumn{1}{c|}{\multirow{2}{*}{\textbf{MODEL}}} & \multicolumn{5}{c|}{\textbf{F1-SCORE}} & \multicolumn{5}{c|}{\textbf{PRECISION}} & \multicolumn{5}{c}{\textbf{RECALL}} \\
\multicolumn{1}{c|}{} & AVG & BR & NR & OT & SA & AVG & BR & NR & OT & SA & AVG & BR & NR & OT & SA \\
\midrule
mistral-7b & .703 & .761 & .680 & .785 & .587 & .732 & \textbf{\underline{.766}} & \textbf{\underline{.665}} & .669 & .827 & .714 & .755 & .695 & \textbf{\underline{.950}} & .455 \\
llama-3b & .257 & .395 & .319 & .204 & .109 & .459 & .300 & .249 & .686 & .600 & .301 & .580 & .445 & .120 & .060 \\
gemma-9b & \textbf{\underline{.716}} & \textbf{\underline{.763}} & \textbf{\underline{.715}} & \textbf{\underline{.787}} & \textbf{\underline{.597}} & \textbf{\underline{.775}} & .662 & .633 & .829 & \textbf{\underline{.977}} & \textbf{\underline{.725}} & \textbf{\underline{.900}} & \textbf{\underline{.820}} & .750 & .430 \\
phi-3 & .576 & .627 & .577 & .675 & .423 & .702 & .518 & .473 & \textbf{\underline{.900}} & .917 & .588 & .795 & .740 & .540 & .275 \\
zephyr-7b & .486 & .448 & .407 & .530 & .557 & .516 & .607 & .496 & .429 & .532 & .495 & .355 & .345 & .695 & \textbf{\underline{.585}} \\
\midrule
AVG & .548 & \textbf{\underline{.599}} & .540 & .596 & .455 & .637 & .571 & .503 & .703 & \textbf{\underline{.771}} & .565 & \textbf{\underline{.677}} & .609 & .611 & .361 \\
\bottomrule
\end{tabular}%
}
\end{table}

\subsection{Fine-Tuning}\label{results-fine-tuning}

Fine-tuning on the \textit{MeSH-Rel-4K} substantially improved performance across all models compared to prompting strategies, as reported in Table~\ref{table:FT-MeSH-4K}.
When evaluated against the CoT, two-way approach, fine-tuning yielded a significant average F1-score improvement of 34.1 percentage points (61.6\% relative performance increase).
While \texttt{gemma-9b} retained overall dominance (F1: 91.6\%), the smallest model (\texttt{llama-3b}) demonstrated an improvement of 60.5 percentage point (from F1: 25.7\% with CoT, two-way to F1: 86.2\% after fine-tuning). Nevertheless, other models exhibit distinct, metric-specific strengths. For instance, \texttt{mistral-7b} achieves the highest precision for the \texttt{other} category (96.8\%), whilst \texttt{zephyr-7b} attains the highest recall for \texttt{narrower} (88.0\%), and \texttt{llama-3b} for \texttt{broader} (95.0\%).


Across all evaluated models, the four categories achieve comparable average F1-scores ranging from 86.6\% to 94.3\%. The \texttt{other} category consistently yields the highest average scores across all metrics (F1: 94.3\%, precision: 94.9\%, and recall: 93.8\%). These exceptionally high results demonstrate that fine-tuned LLMs efficiently identify and isolate unrelated topics. This is important for ontology generation, as incorrectly hallucinated relations can compromise the structural integrity of an ontology by introducing loops~\cite{osborne2015klink}.
Conversely, \texttt{same-as} proves to be the most challenging relation, recording the lowest average F1-score (86.3\%) and precision (87.3\%) across the cohort. Followed by \texttt{narrower} exhibiting the lowest average recall (84.9\%). 
These findings confirm that fine-tuning effectively bridges the reasoning deficits of smaller models, yielding near-state-of-the-art models.

\begin{table}[!h]
\centering
\caption{Precision, Recall, and F1-score for models fine-tuned and evaluated on the \textit{MeSH-Rel-4K}. BR refers to performance on \texttt{broader} relations, NR to \texttt{narrower}, OT to \texttt{other}, and SA to \texttt{same-as}. AVG indicates the average performance across the four categories. The best-performing scores for each relation are highlighted in {\underline{\textbf{bold \& underlined}}}. Due to space constraints, the leading zero has been omitted from all values.}
\label{table:FT-MeSH-4K}
\resizebox{\columnwidth}{!}{%
\begin{tabular}{l|ccccc|ccccc|ccccc}
\toprule
\multicolumn{1}{c|}{\multirow{2}{*}{\textbf{MODEL}}} & \multicolumn{5}{c|}{\textbf{F1-SCORE}} & \multicolumn{5}{c|}{\textbf{PRECISION}} & \multicolumn{5}{c}{\textbf{RECALL}} \\
\multicolumn{1}{c|}{} & AVG & BR & NR & OT & SA & AVG & BR & NR & OT & SA & AVG & BR & NR & OT & SA \\
\midrule
mistral-7b & .888 & .880 & .871 & .941 & .860 & .889 & .847 & .882 & \textbf{\underline{.968}} & .860 & .887 & .915 & .860 & .915 & .860 \\
llama-3b & .862 & .858 & .833 & .927 & .828 & .868 & .782 & .857 & .967 & .864 & .861 & \textbf{\underline{.950}} & .810 & .890 & .795 \\
gemma-9b & \textbf{\underline{.916}} & \textbf{\underline{.923}} & \textbf{\underline{.888}} & \textbf{\underline{.958}} & \textbf{\underline{.895}} & \textbf{\underline{.916}} & \textbf{\underline{.920}} & \textbf{\underline{.906}} & .946 & \textbf{\underline{.891}} & \textbf{\underline{.916}} & .925 & .870 & \textbf{\underline{.970}} & \textbf{\underline{.900}} \\
phi-3 & .877 & .854 & .857 & .941 & .856 & .878 & .830 & .892 & .923 & .867 & .877 & .880 & .825 & .960 & .845 \\
zephyr-7b & .900 & .895 & .880 & .948 & .877 & .900 & .895 & .880 & .941 & .883 & .900 & .895 & \textbf{\underline{.880}} & .955 & .870 \\
\midrule
AVG & .889 & .882 & .866 & \textbf{\underline{.943}} & .863 & .890 & .855 & .883 & \textbf{\underline{.949}} & .873 & .888 & .913 & .849 & \textbf{\underline{.938}} & .854 \\
\bottomrule
\end{tabular}%
}
\end{table}

\subsection{Analysis of LLMs on MeSH-Rel-4K}\label{results-mesh-4k}

\begin{figure*}[t!]
    \centering
    \begin{subfigure}[t]{0.50\textwidth}
        \centering
        \includegraphics[width=\linewidth]{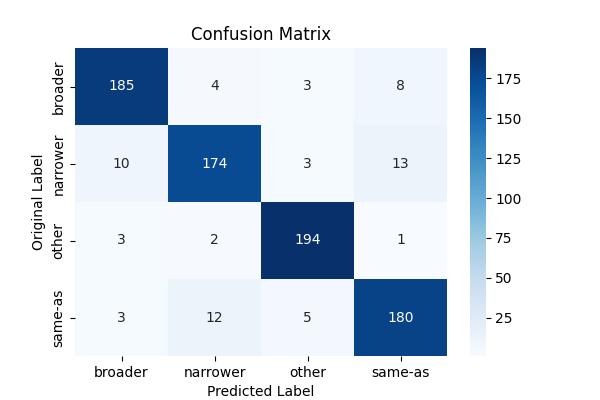}
        \caption{\texttt{gemma-9b} (F1: 91.60\%)}\label{gemma-9b}
    \end{subfigure}%
    \begin{subfigure}[t]{0.50\textwidth}
        \centering
        \includegraphics[width=\linewidth]{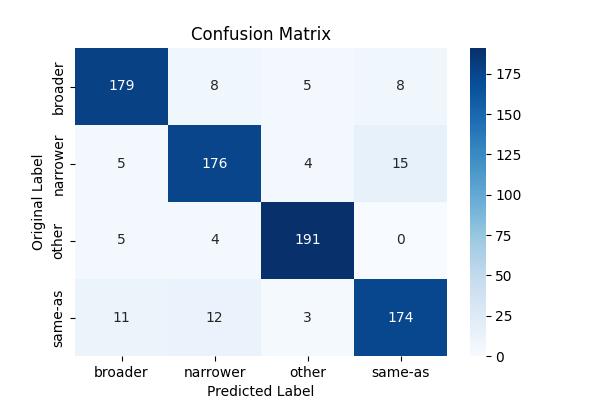}
        \caption{\texttt{zephyr-7b} (F1: 90.00\%)}\label{zephyr-7b}
    \end{subfigure}
    \begin{subfigure}[t]{0.50\textwidth}
        \centering
        \includegraphics[width=\linewidth]{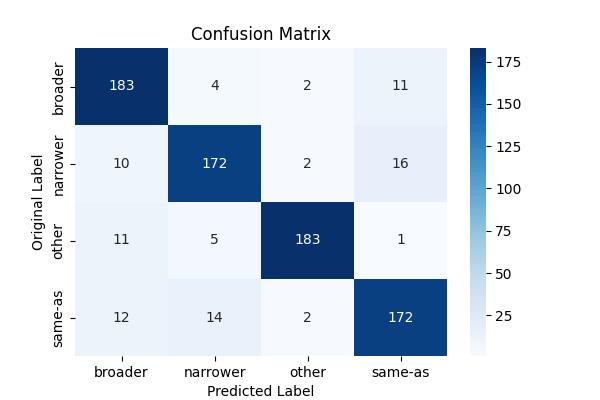}
        \caption{\texttt{mistral-7b} (F1: 88.80\%) }\label{mistral-7b}
    \end{subfigure}%
    \begin{subfigure}[t]{0.50\textwidth}
        \centering
        \includegraphics[width=\linewidth]{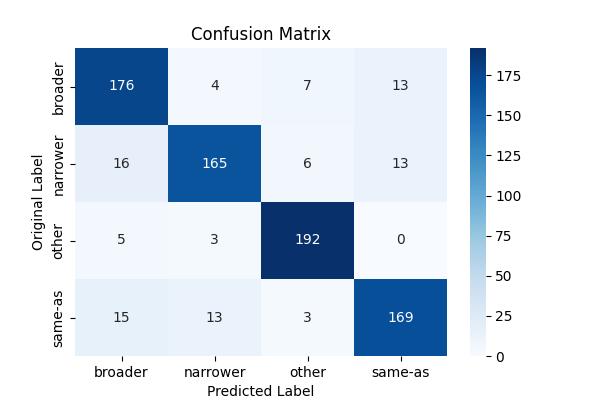}
        \caption{\texttt{phi-3} (F1: 87.70\%)}\label{phi-3}
    \end{subfigure}
    \begin{subfigure}[t]{0.50\textwidth}
        \centering
        \includegraphics[width=\linewidth]{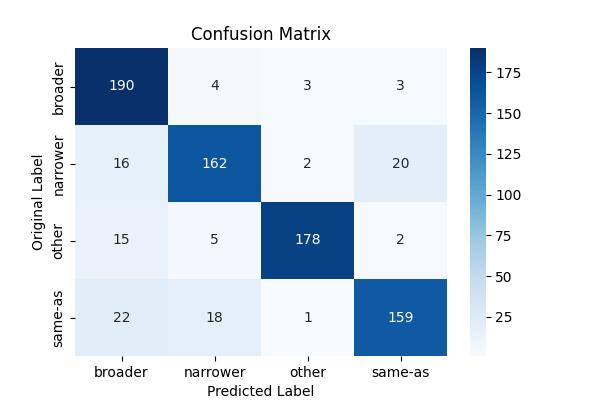}
        \caption{\texttt{llama-3b} (F1: 86.20\%)}\label{llama-3b}
    \end{subfigure}
    \caption{Confusion matrices for LLMs fine-tuned and evaluated on \textit{MeSH-Rel-4K}}
    \label{fig:confusionmatrices}
\end{figure*}

The confusion matrices, presented in Fig.~\ref{fig:confusionmatrices}, indicate that all fine-tuned models achieve high classification accuracy, with F1-scores ranging from 86.2\% to 91.6\%.
The \texttt{other} category consistently yields the highest true positive rates across the cohort (194 out of 200 instances for \texttt{gemma-9b}, see Fig.~\ref{gemma-9b}). This demonstrates that differentiating unrelated topics presents minimal difficulty for fine-tuned LLMs.
The primary classification bottleneck across all models is the disambiguation of equivalence (\texttt{same-as}) from hierarchical relations (\texttt{broader}/\texttt{narrower}). The frequency of these boundary errors scales inversely with model parameter count. While \texttt{gemma-9b} minimises this confusion, the smallest model (\texttt{llama-3b}, see Fig.~\ref{llama-3b}) frequently predicts false hierarchical relations, misclassifying 40 \texttt{same-as} instances as \texttt{broader} (22) and \texttt{narrower} (18). 
The mid-sized models (\texttt{zephyr-7b}, \texttt{mistral-7b}, and \texttt{phi-3}) exhibit symmetric confusion, misclassifying hierarchical and equivalence relationships at comparable rates. For instance, \texttt{zephyr-7b} (see Fig.~\ref{zephyr-7b}) misclassifies 23 hierarchical instances (8 \texttt{broader}, 15 \texttt{narrower}) as \texttt{same-as}, whilst predicting exactly 23 \texttt{same-as} instances as \texttt{broader} (11) and \texttt{narrower} (12). This bidirectional error distribution is mirrored by \texttt{mistral-7b} (see Fig.~\ref{mistral-7b}), which predicts 27 hierarchical topics (11 \texttt{broader}, 16 \texttt{narrower}) as \texttt{same-as} against 26 \texttt{same-as} instances classified as \texttt{broader} (12) and \texttt{narrower} (14). \texttt{phi-3} (see Fig.~\ref{phi-3}) replicates this pattern, misattributing 26 hierarchical instances (13 \texttt{broader}, 13 \texttt{narrower}) to \texttt{same-as}, alongside 28 \texttt{same-as} instances to hierarchical categories (15 \texttt{broader}, 13 \texttt{narrower}).

Analysis of these misclassifications indicates that this error pattern correlates with dense lexical overlap within the MeSH. For example, ``alitretinoin'' (9-cis-retinoic acid) is formally defined as a \texttt{narrower} topic relative to the foundational compound ``tretinoin''. However, all five fine-tuned models unanimously classified this relationship as \texttt{same-as}.

\section{Conclusion and Future Work}\label{conclusion}

This paper evaluated the performance of five small, open-source LLMs in identifying semantic relationships between research topics within the biomedical domain. To support this analysis, we introduced \textit{MeSH-Rel-4K} (comprising 4K semantic relationships) extracted from the MeSH.
Our experiments demonstrated that while advanced prompting strategies (such as CoT, two-way~\cite{aggarwal2026large}) yielded moderate baseline improvements, explicit fine-tuning on \textit{MeSH-Rel-4K} significantly increased classification accuracy. The fine-tuned \texttt{gemma-9b} achieved the highest overall performance (F1: 91.6\%). Furthermore, \texttt{llama-3b} exhibited a 60.5 percentage point increase in its F1-score following fine-tuning.

Our future work will focus on broadening our analysis across various academic disciplines. Beginning with STEM fields, we will expand into the Humanities, Social Sciences, and Economics. Additionally, we plan to develop a comprehensive pipeline for ontology matching and evolution.
To identify semantic links between existing ontologies, this pipeline will use an LLM-based classifier. It will extract relevant metadata\footnote{Metadata: Title, Abstract, Keywords, Introduction, and Conclusion} from scholarly articles and identify emerging research topics. These topics will be systematically integrated into a self-evolving, multidisciplinary ontology of research topics.


\section*{Declaration on Generative AI}
During the preparation of this work, the author(s) used Grammarly, and Gemini to check grammar and spelling. After using this tool/service, the author(s) reviewed and edited the content as needed and take(s) full responsibility for the content of the publication. 

\bibliography{sample-ceur}


\end{document}